\documentclass[prc,preprint]{revtex4}
\usepackage{graphicx}
\usepackage{subfigure}
\usepackage{amsmath,amssymb}
\usepackage{mathrsfs}
\usepackage{latexsym}
\usepackage{graphics}
\usepackage{amsmath,amssymb}
\usepackage{mathrsfs}
\usepackage[english]{babel}
\usepackage{setspace}
\usepackage{ctable}

\begin{document}
\def\Journal#1#2#3#4{{#1} {\bf{#2}}, {#3} (#4).}
\def\ANP{Adv. Nucl. Phys.}
\def\ARNPS{Ann. Rev. Nucl. Part. Sci.}
\def\CTP{Commun. Theor. Phys.}
\def\EPJA{Eur. Phys. J. A}
\def\EPJC{Eur. Phys. J. C}
\def\IJMPA{International Journal of Modern Physics A}
\def\IJMPE{International Journal of Modern Physics E}
\def\JCHP{J. Chem. Phys.}
\def\JCP{Journal of Computational Physics}
\def\JHEP{JHEP}
\def\JPCS{Journal of Physics: Conference Series}
\def\JPG{J. Phys. G: Nucl. Part. Phys.}
\def\NATURE{Nature}
\def\NC{La Rivista del Nuovo Cimento}
\def\NCA{IL Nuovo Cimento A}
\def\NPA{Nucl. Phys. A}
\def\NST{Nuclear Science and Techniques}
\def\PA{Physica A}
\def\PAN{Physics of Atomic Nuclei}
\def\PHY{Physics}
\def\PRA{Phys. Rev. A}
\def\PRC{Phys. Rev. C}
\def\PRD{Phys. Rev. D}
\def\PLA{Phys. Lett. A}
\def\PLB{Phys. Lett. B}
\def\PLD{Phys. Lett. D}
\def\PRL{Phys. Rev. Lett.}
\def\PL{Phys. Lett.}
\def\PREV{Phys. Rev.}
\def\PREP{\em Physics Reports}
\def\PROG{Progress in Particle and Nuclear Physics}
\def\RPP{Rep. Prog. Phys.}
\def\RDNC{Rivista del Nuovo Cimento}
\def\RMP{Rev. Mod. Phys}
\def\SCIENCE{Science}
\def\ZPA{Z. Phys. A.}

\def\ANN{Ann. Rev. Nucl. Part. Sci.}
\def\ANNAST{Ann. Rev. Astron. Astrophys.}
\def\AP{Ann. Phys}
\def\APJ{Astrophysical Journal}
\def\APJS{Astrophys. J. Suppl. Ser.}
\def\EJP{Eur. J. Phys.}
\def\LANC{Lettere Al Nuovo Cimento}
\def\NCA{Nuovo Cimento A}
\def\PHYS{Physica}
\def\NP{Nucl. Phys}
\def\MATH{J. Math. Phys.}
\def\JPAM{J. Phys. A: Math. Gen.}
\def\PRO{Prog. Theor. Phys.}
\def\NPB{Nucl. Phys. B}

\title{Systematic analysis of hadron spectra in p+p collisions using Tsallis distribution}
\author{H. Zheng$^{1,2, 3}$, Lilin Zhu$^{4}$,  A. Bonasera $^{1, 2}$
}                     
%
%
\affiliation{$^1$Laboratori Nazionali del Sud, INFN, via Santa Sofia, 62, 95123 Catania, Italy\\
$^2$Cyclotron Institute, Texas A\&M University, College Station, TX 77843, USA \\
$^3$Physics Department, Texas A\&M University, College Station, TX 77843, USA\\ 
$^4$Department of Physics, Sichuan University, Chengdu 610064, P. R. China 
}
%
%

\begin{abstract}
Using the experimental data from the  STAR, PHENIX, ALICE and CMS programs on the rapidity and energy dependence of the $p_T$ spectra in p+p collisions, we show that a universal distribution exists. The energy dependence of temperature $T$ and parameter $n$ of the Tsallis distribution are also discussed in detail. A cascade particle production mechanism in p+p collisions is proposed.

\end{abstract}
\maketitle
\section{Introduction}
The particle spectrum is a basic quantity measured in experiments and it can reveal the information of particle production mechanism in heavy-ion collisions. Recently, the Tsallis distribution has attracted many theorists' and experimentalists' attention in high energy heavy-ion collisions \cite{star2007, phenix2011, alice1, alice2, aliceS2012, cms3, cmsdata900, cmsdata7000, cms1, sena,  liuAuAu2014, khandaiflow,  cms2014,  cleymans0, cleymans, azmiJPG2014,  liAuAu2013, maciej, khandai, wongprd, wong2012, wongarxiv2014, maciej1,cleymans2}. The excellent ability to fit the spectra of identified hadrons and charged particles in a large range of $p_T$ up to 200 GeV is quite impressive \cite{wongprd, wong2012, wongarxiv2014, maciej1}. From the phenomenological view, there may be real physics behind the prominent phenomenology work, e.g. Regge trajectory for particle classification \cite{wongbook}. p+p collision experiments have been performed and measured under different energies. Since p+p collision is very simple compared to nucleus-nucleus collision, the measurements of p+p collisions are used to understand the particle interaction, particle production mechanism and as a baseline for nucleus-nucleus collisions. Many efforts have been put to study the particle spectra produced in p+p collisions using Tsallis distribution. Different versions of the Tsallis distribution are used in the literature \cite{star2007, phenix2011, alice1, alice2, aliceS2012, cms3, cmsdata900, cmsdata7000, cms1, sena,  liuAuAu2014,khandaiflow,  cms2014, cleymans0, cleymans, azmiJPG2014,  liAuAu2013, maciej, khandai, wongprd, wong2012, wongarxiv2014, maciej1,cleymans2}. The parameter $T$ in the Tsallis distribution was interpreted as temperature. All of them can fit the particle spectra very well but they give different temperatures. In this work, we would like to study the connections and differences among different versions of the Tsallis distribution. We collected p+p collisions data with different $p_T$ ranges and different rapidity cuts from different experiment groups at RHIC and LHC  and did a systematic study of the particle spectra using one of the Tsallis distributions. 

The paper is organized as following. In section II, we review different versions of the Tsallis distribution used in the literature and analyze their connections and differences. We also give the form of Tsallis distribution used in our anlysis. In section III, we show our results of particle spectra from p+p collisions and analyze them. We compare our results with the ones obtained by the other authors in the literature. A brief conclusion is given in the section IV.  
\label{intro}

\section{Tsallis distributions}
The STAR \cite{star2007}, PHENIX \cite{phenix2011} Collaborations at RHIC and ALICE \cite{alice1, alice2, aliceS2012} and CMS \cite{cms3} Collaborations at LHC adopted the form of Tsallis distribution
\begin{eqnarray}
E\frac{d^3N}{dp^3}&=&\frac{1}{2\pi p_T} \frac{d^2N}{dydp_T} \nonumber\\ 
&=& \frac{dN}{dy} \frac{(n-1)(n-2)}{2\pi nC[nC+m(n-2))]}(1+\frac{m_T-m}{nC})^{-n}, \label{exptsallis}
\end{eqnarray}
where $m_T=\sqrt{p_T^2+m^2}$ is the transverse mass. $m$ was used as a fitting parameter in ref. \cite{star2007}, but it represents the rest mass of the particle studied in refs. \cite{phenix2011, alice1, alice2, aliceS2012, cms3}.  $\frac{dN}{dy}$, $n$ and $C$ are fitting parameters. With this form of the Tsallis distribution, when $p_T\gg m$, we can ignore the $m$ in the last term in Eq. (\ref{exptsallis}) and obtain $E\frac{d^3N}{dp^3} \propto p_T^{-n}$. This result is well known because high energy particles come from hard scattering and they follow a power law distribution with $p_T$. When $p_T\ll m$ which is the non-relativistic limit, we obtain $m_T -m=\frac{p_T^2}{2m}=E_{T}^{classical}$ and $E\frac{d^3N}{dp^3} \propto e^{-\frac{E_T^{classical}}{C}}$, i.e. a thermal distribution. The parameter $C$ in Eq. (\ref{exptsallis}) plays the same role as temperature $T$.

In refs. \cite{cleymans0, cleymans, azmiJPG2014, liAuAu2013, maciej}, the following Tsallis form  is used
\begin{eqnarray}
E\frac{d^3N}{dp^3} = gV\frac{m_T \cosh y}{(2\pi)^3} [1+(q-1)\frac{m_T\cosh y-\mu}{T}]^{\frac{q}{1-q}}, \label{tsallisB}
\end{eqnarray}
based on thermodynamic consistency arguments. Where $g$ is the degeneracy of the particle, $V$ is the volume, $y$ is the rapidity, $\mu$ is the chemical potential, $T$ is the temperature and $q$ is the entropic factor, which measures the nonadditivity of the entropy. In Eq. (\ref{tsallisB}), there are four parameters $V, \mu, T, q$. $\mu$ was assumed to be 0 in refs. \cite{cleymans0, cleymans, azmiJPG2014, liAuAu2013} which is a reasonable assumption because the energy is high enough and the chemical potential is small compared to temperature. In the mid-rapidity $y=0$ region, Eq. (\ref{tsallisB}) is reduced to
\begin{equation}
E\frac{d^3N}{dp^3} = gV\frac{m_T}{(2\pi)^3} [1+(q-1)\frac{m_T}{T}]^{q/(1-q)}. \label{tsallisBR}
\end{equation}
It becomes very similar to Eq. (\ref{exptsallis}), but there are some differences, i.e. $m_T$ replaces $m_T-m$ in the bracket and there is a term $m_T$ in front of the bracket as well. We have seen that two kinds of representation have been used: one is with parameter $n$, i.e. Eq. (\ref{exptsallis}) and the other is with parameter $q$, i.e. Eq. (\ref{tsallisBR}). There is no direct match between $n$ and $q$ because Eqs. (\ref{exptsallis}, \ref{tsallisBR}) are just similar. We find a connection between $n$ and $q$ in the limit at large $p_T$. 


When $p_T\gg m$, from Eq. (\ref{tsallisBR}), we can obtain
\begin{equation}
E\frac{d^3N}{dp^3} \propto p_T^{-\frac{1}{q-1}}. \label{tsallisBRlimit}
\end{equation}
Recalling that $E\frac{d^3N}{dp^3}\propto p_T^{-n}$ when $p_T\gg m$ from Eq. (\ref{exptsallis}), therefore the relation between $n$ and $q$ is
\begin{equation}
n = \frac{1}{q-1}. \label{nq}
\end{equation}
Another treatment to find the relation between $n$ and $q$ can be found in ref. {\cite{cleymans2}}.

In ref. \cite{sena}, Sena {\it et al.} applied the non-extensive formalism to obtain the probability of particle with mometum $p_T$
\begin{equation}
\frac{1}{\sigma}\frac{d\sigma}{dp_T} = c  p_T\int_0^\infty dp_L [1+(q-1)\beta \sqrt{p_L^2+p_T^2 + m^2}]^{-q/(q-1)},\label{senaeq}
\end{equation}
where $c$ is the normalization constant, $q$ is a parameter, $\beta=\frac{1}{T}$ and $m$ is the mass of particle.
With the approximation $p_T$ very large compared to $p_L$ and $m$ \cite{beck}, Eq. (\ref{senaeq}) can be rewritten as
\begin{eqnarray}
\frac{1}{\sigma}\frac{d\sigma}{dp_T} &=& c[2(q-1)]^{-1/2}B(\frac{1}{2}, \frac{q}{q-1}-\frac{1}{2})u^{3/2}[1+(q-1)u]^{-\frac{q}{q-1}+\frac{1}{2}},\label{tsallislong}
\end{eqnarray}
where $u=\frac{p_T}{T}$ and B(x, y) is the Beta-function. We can see that Eq. (\ref{tsallislong}) is also similar to Eqs. (\ref{exptsallis}, \ref{tsallisBR}) but not exactly the same. We repeat the same limit condition, let $p_T$ be very large, $q\ne 1$ and $(q-1)u\gg 1$, then
\begin{equation}
\frac{1}{\sigma}\frac{d\sigma}{p_Tdp_T} \propto p_T^{-\frac{1}{q-1}},
\end{equation}
which is the same as Eq. (\ref{tsallisBRlimit}).

In the previous versions of the Tsallis distribution, the rapidity cuts for the experimental data are not taken into account. In ref. \cite{wong2012}, Wong {\it et al.} proposed a new form of the Tsallis distribution function to take into account the rapidity cut,
\begin{equation}
(E\frac{d^3N}{dp^3})_{|\eta|<a}=\int_{-a}^{a}d\eta \frac{dy}{d\eta}(\frac{d^3N}{dp^3}). \label{tsalliswong}
\end{equation}
Where
\begin{eqnarray}
\frac{dy}{d\eta}(\eta, p_T)=\sqrt{1-\frac{m^2}{m^2_T\cosh^2 y}}, 
\end{eqnarray}
with
$$y=\frac{1}{2}\ln \Big [\frac{\sqrt{p_T^2 \cosh^2 \eta + m^2}+p_T\sinh \eta}{\sqrt{p_T^2 \cosh^2 \eta + m^2}-p_T\sinh \eta}\Big],$$
and
\begin{equation}
\frac{d^3N}{dp^3}=C\frac{dN}{dy}(1+\frac{E_T}{nT})^{-n},\quad 
E_T=m_T-m, \label{tsalliswongdndp}
\end{equation}
where $C\frac{dN}{dy}$ is assumed to be a constant in ref. \cite{wong2012}.

\label{sec:1}
\begin{figure}
\resizebox{0.5\textwidth}{!}{%
  \includegraphics{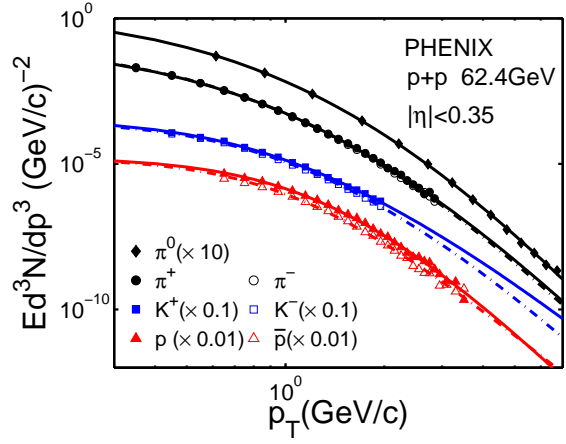}
}
\caption{(Color online) Fitting results using the Tsallis distribution Eq. (\ref{tsallisus}) for $\pi^0$, $\pi^{\pm}$, $K^{\pm}$, $p$ and $\bar p$ in p+p collisions at $\sqrt{s}=62.4$ GeV. Data are taken from PHENIX \cite{phenix2011, phenix2009}.}
\label{Fig1}       
\end{figure}

We can rewrite Eq. (\ref{tsalliswongdndp}) into
\begin{equation}
\frac{d^3N}{dp^3}=C\frac{dN}{d\eta} \frac{d\eta}{dy}(1+\frac{E_T}{nT})^{-n}. \label{tsalliswongdndp1}
\end{equation}
From the experimental measurements, we know that $\frac{dN}{d\eta}$ is almost a constant when $\eta$ is not large \cite{dndeta}. 
We can simply treat it as a constant. Subsituting Eq. (\ref{tsalliswongdndp1}) into Eq. (\ref{tsalliswong}), we obtain
\begin{equation}
(E\frac{d^3N}{dp^3})_{|\eta|<a} = A(1+\frac{E_T}{nT})^{-n}, \label{tsallisus}
\end{equation}
where all the constants are absorbed into the new fitting parameter $A$. $n$ and $T$ are the other two fitting parameters. This form of Tsallis distribution was derived without resorting to thermodynamical description, which is different from Eq. (2). This also causes the power differences in Eqs. (\ref{tsallisB}) and (\ref{tsallisus}), i.e. the power $q/(1-q) = -nq$ in Eq. (\ref{tsallisB}). Eq. (\ref{tsallisus}) is equivalent to Eq. (\ref{exptsallis}) but in a simpler form. We adopt it to do the systematic particle spectra analysis from p+p collisions. We will show that the approximation in Eq. (\ref{tsallisus}) doesn't change the results from Eq. (\ref{tsalliswong}) later in this paper. We notice that Eq. (\ref{tsallisus}) has been used by CMS Collaboration \cite{cmsdata900, cmsdata7000, cmsB2013} and by Wong {\it et al.} in their recent paper \cite{wongarxiv2014}. The STAR Collaboration also applied a formula which is very close to Eq. (\ref{tsallisus}) \cite{star2010}.

\begin{figure*}
        \begin{tabular}{ccc}
        \includegraphics[width=0.35\textwidth]{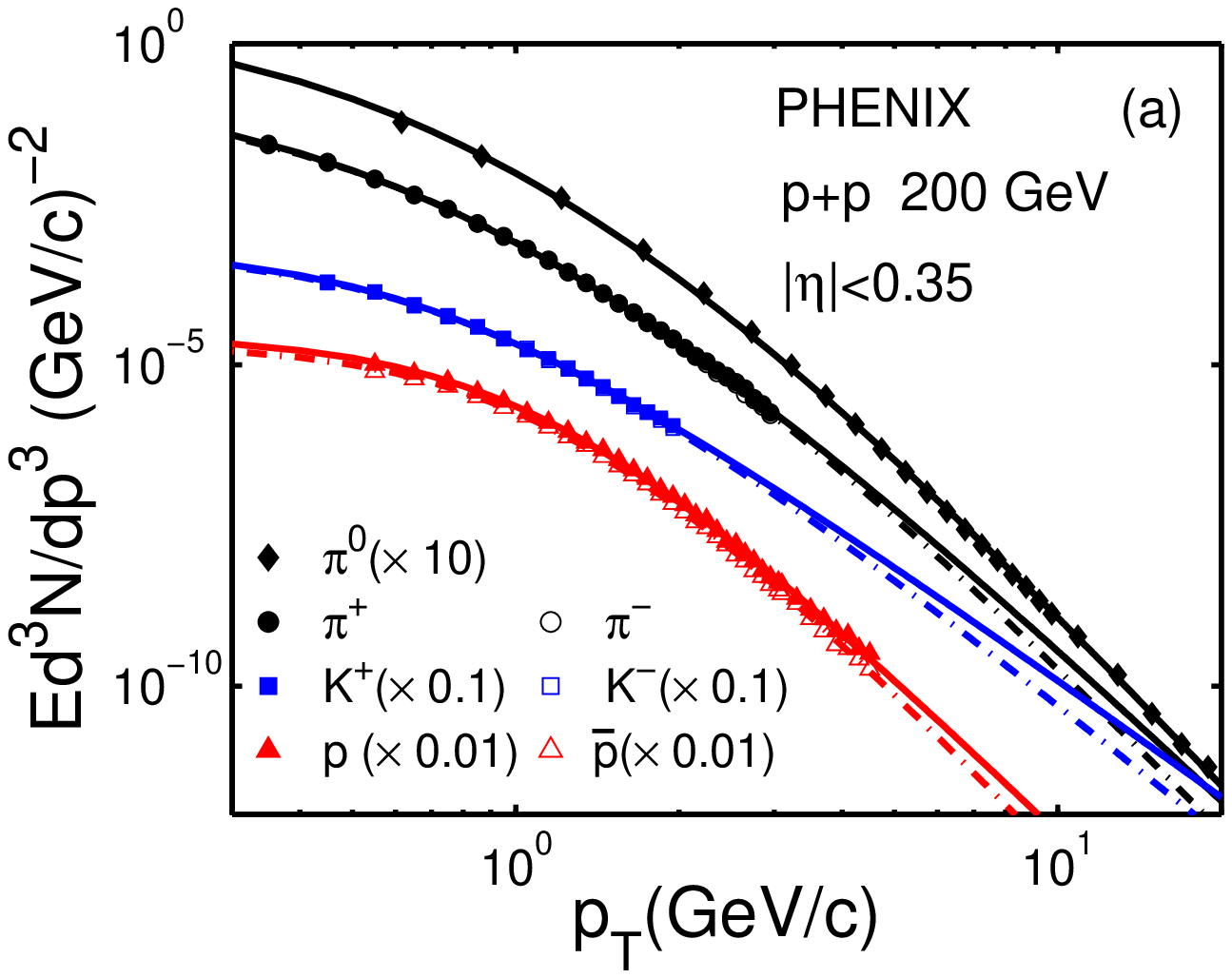}  \includegraphics[width=0.35\textwidth]{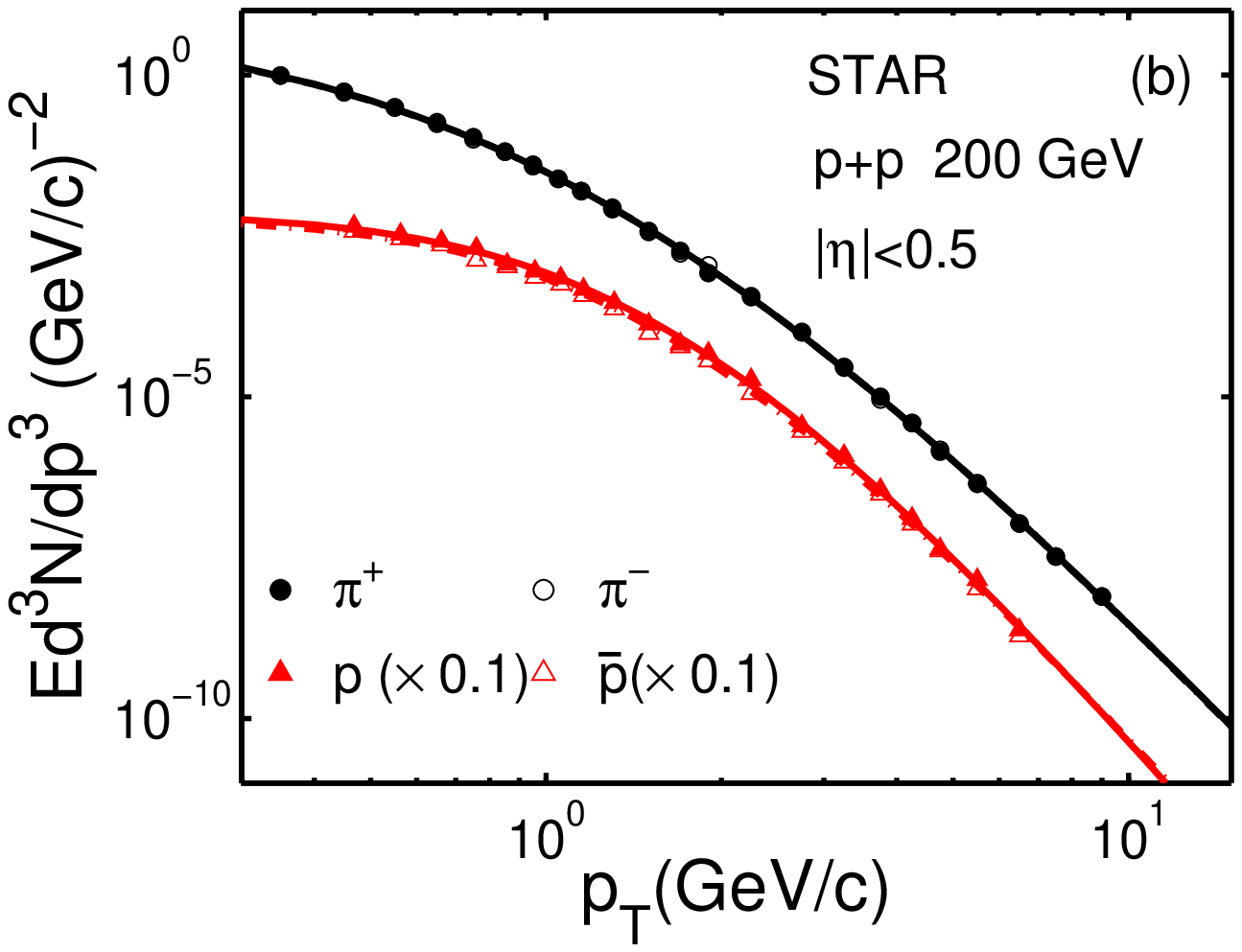}
        \includegraphics[width=0.35\textwidth]{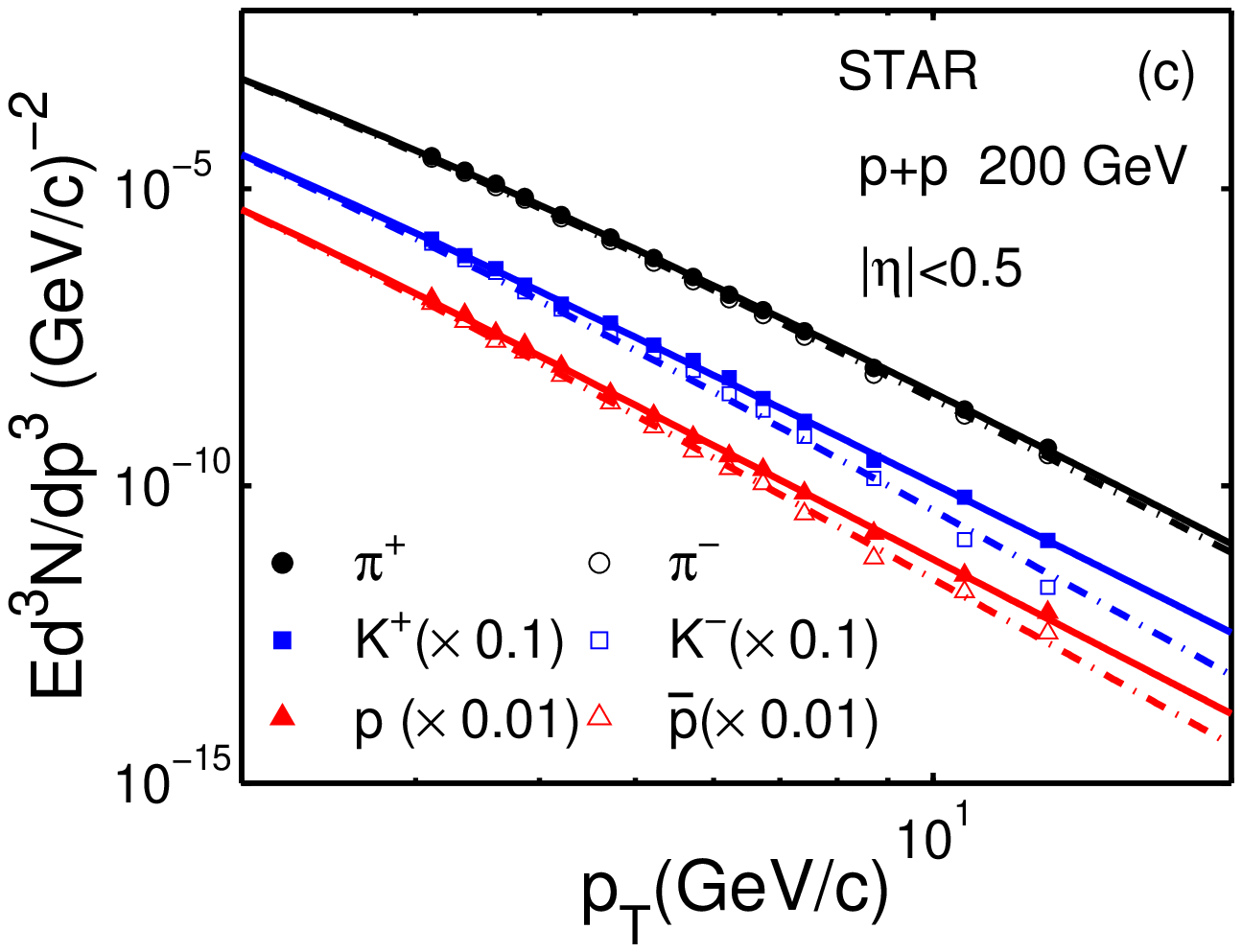}
               \end{tabular}
\caption{(Color online) Fitting results using the Tsallis distribution Eq. (\ref{tsallisus}) for $\pi^0$, $\pi^{\pm}$, $K^{\pm}$, $p$ and $\bar p$ in p+p collisions at $\sqrt{s}=200$ GeV. The data in (a) are taken from PHENIX \cite{phenix2011, phenix2007} and the data in (b) and (c) are taken from STAR but have different $p_T$ ranges  \cite{star2, star1}. }
\label{Fig2}       
\end{figure*}


        \begin{figure*}  
        \centering
        \begin{tabular}{cc}
        \includegraphics[width=0.45\textwidth]{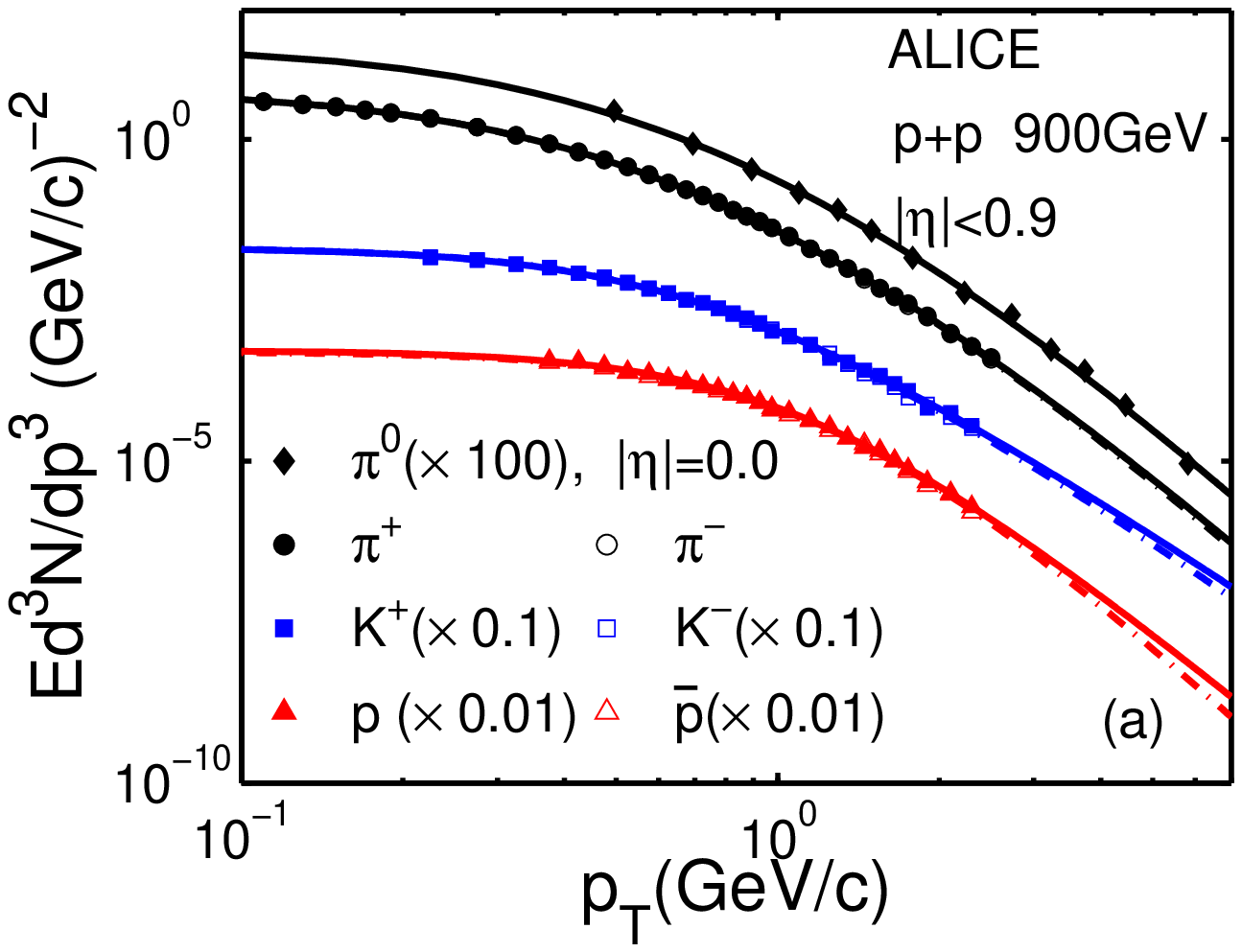}
        \includegraphics[width=0.45\textwidth]{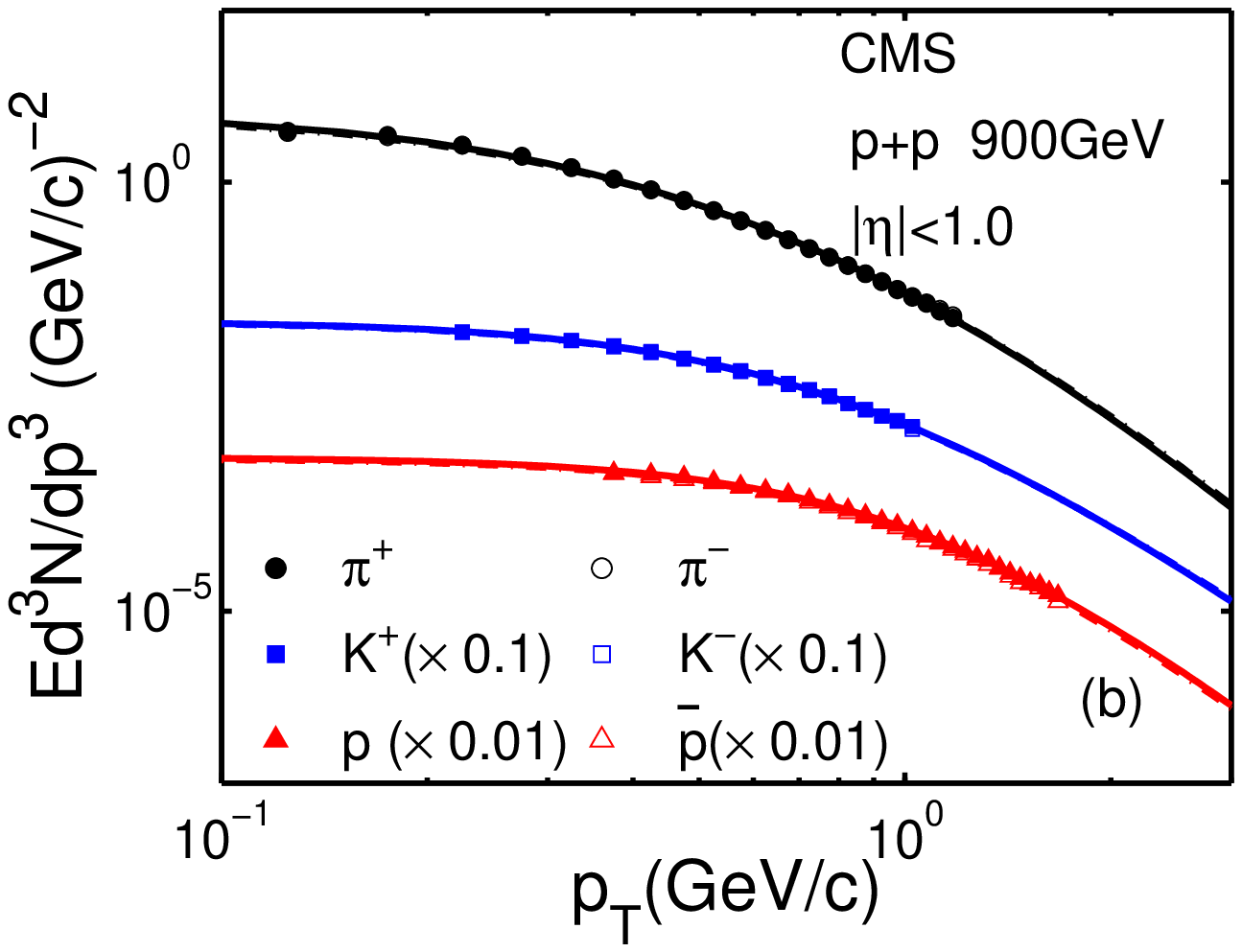}
               \end{tabular}
\caption{(Color online) Fitting results using the Tsallis distribution Eq. (\ref{tsallisus}) for $\pi^0$, $\pi^{\pm}$, $K^{\pm}$, $p$ and $\bar p$ in p+p collisions at $\sqrt{s}=900$ GeV. Data are taken from (a) ALICE \cite{alice1, alice2} and (b) CMS \cite{cms3}.} \label{Fig3}
    \end{figure*}
 
    
     \begin{figure}  
\resizebox{0.5\textwidth}{!}{%
  \includegraphics{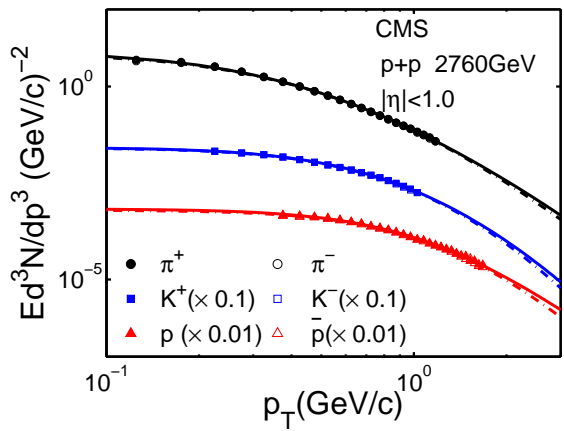}
}
\caption{(Color online) Fitting results using the Tsallis distribution Eq. (\ref{tsallisus}) for $\pi^{\pm}$, $K^{\pm}$, $p$ and $\bar p$ in p+p collisions at $\sqrt{s}=2760$ GeV. Data are taken from CMS \cite{cms3}.} \label{Fig4}
    \end{figure}   
    

\section{Results}
\label{sec:2}
We collect the spectra data for different particles with different $p_T$ ranges and different rapidity cuts from p+p collisions at $\sqrt{s} = 62.4, 200, 900, 2760$ and $7000$ GeV. A similar study has been done in ref. \cite{khandai} assuming all the particles have the same temperatures while for the pions, which are produced the most in p+p collisions, the temperature is a free parameter. It is not necessary to require that the temperatures for all the particles are the same. If all the particles are produced at the same time and in thermal equilibrium, it is reasonable to have this constraint. If, for instance, the particles are produced at different times, i.e. in the framework of a cascade particle production mechanism, the particles will not have the same temperatures because of energy conservation. The particles produced at early time will have the higher temperature than the ones produced at later time. In this analysis, we allow the temperature to be a free parameter for different particles. This work is the natural extension of ref. \cite{khandai}. Meanwhile, we include more data and a wider $p_T$ range. In our analysis, the pion case is exactly the same as in ref. \cite{khandai} except the authors chose another form of the Tsallis distribution, see equation (7) in ref. \cite{khandai}. For the other particles are different. Thus we cannot compare our results with ref. \cite{khandai} directly. We will show this later in the paper. 

        \begin{figure*}  
        \centering
        \begin{tabular}{cc}
        \includegraphics[width=0.45\textwidth]{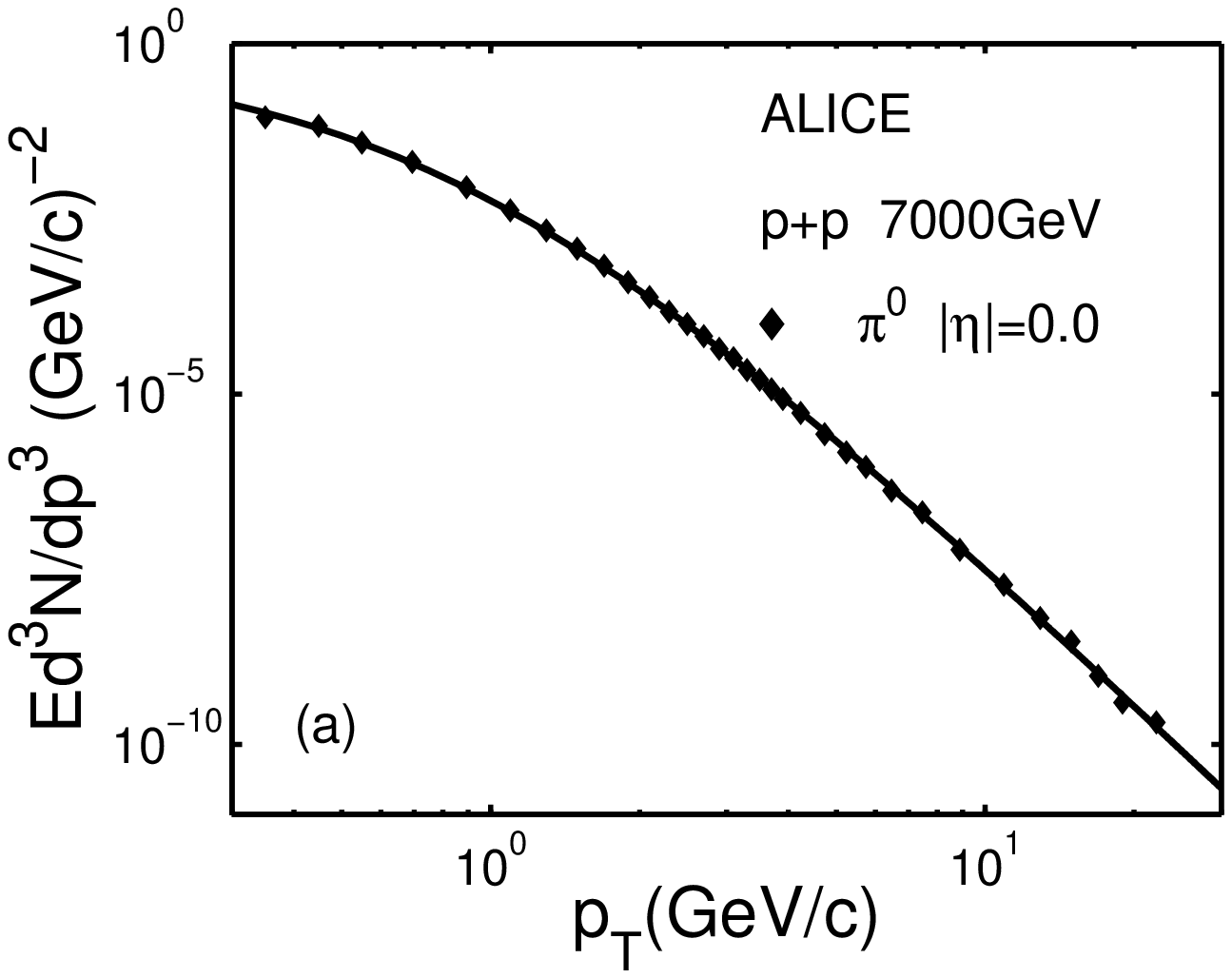}
        \includegraphics[width=0.45\textwidth]{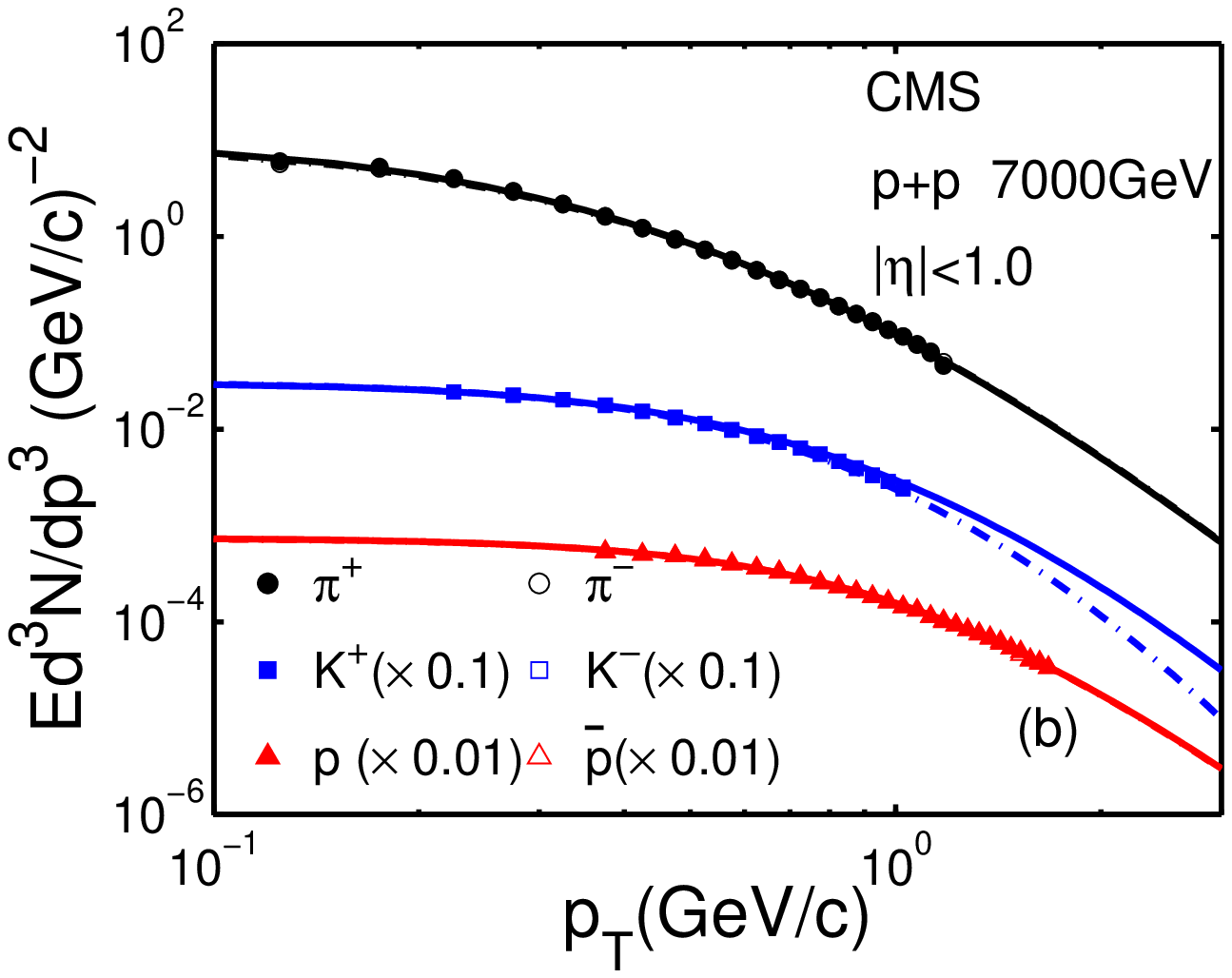}
               \end{tabular}
\caption{(Color online) Fitting results using the Tsallis distribution Eq. (\ref{tsallisus}) for $\pi^0$, $\pi^{\pm}$, $K^{\pm}$, $p$ and $\bar p$ in p+p collisions at $\sqrt{s}=7000$ GeV. Data are taken from (a) ALICE \cite{alice2} and (b) CMS \cite{cms3}. } 
\label{Fig5}
    \end{figure*}
        
     \begin{figure}  
        \centering
\resizebox{0.5\textwidth}{!}{%
  \includegraphics{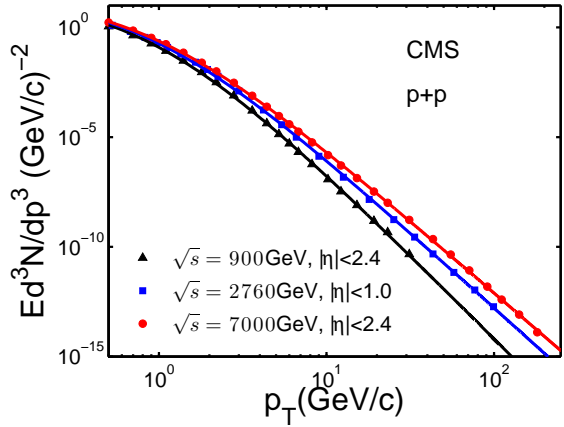}
}
\caption{(Color online) Fitting results using the Tsallis distribution Eq. (\ref{tsallisus}) for charged particles in p+p collisions at $\sqrt{s}=900, 2760$ and 7000 GeV respectively. Data are taken from CMS \cite{cms1, cms2}. }
 \label{Fig6}
    \end{figure} 

In Fig. \ref{Fig1}, we show our fits using the Tsallis distribution for different particles from the PHENIX Collaboration at $\sqrt{s}=62.4$ GeV. We can see that there is only little difference between the mesons and the corresponding anti-particles. The proton spectrum is over that of $\bar p$ because the colliding system is p+p and we have more protons. We can see that our fits with Eq. (\ref{tsallisus}) are good. The fitting parameters can be found in Table. \ref{fitpara}.

Fig. \ref{Fig2} shows the fits for the identified hadrons at $\sqrt{s}=200$ GeV with different $p_T$ ranges and rapidity cuts which are around mid-rapidity.  As we can see, the particle spectra for kaons and protons start to split at high $p_T$. Similar to Fig. \ref{Fig1}, our fits are excellent and the fitting parameters can be found in Table. \ref{fitpara}.

 Similar to Figs. {\ref{Fig1}, \ref{Fig2}}, we repeated the fitting process for different particles produced at higher energies from LHC with different $p_T$ ranges and rapidity cuts. The fitting results are showed in Figs. {\ref{Fig3}, \ref{Fig4}, \ref{Fig5}}. As the collision energy gets higher, there is no difference between the particle and its anti-particle as expected. Same to the fitting results in Figs. {\ref{Fig1}, \ref{Fig2}}, the fitting quality in Figs. {\ref{Fig3}, \ref{Fig4}, \ref{Fig5}} are very good. We list the fitting parameters in Table. \ref{fitpara} as well.
 
  The prominent fitting power of the Tsallis distribution is exhibited perfectly in Fig. \ref{Fig6}. This is for charged particles. As we can see, the excellent fitting can cover 15 orders of magnitude up to 200 GeV/c for $p_T$. This spectacular result was first showed by Wong {\it et al} \cite{wong2012}. In Table. \ref{fitpara} we can see that the fitting results are similar to the pions when we fit the spectra of the identified hadrons separately at the same collision energy. Our fitting results are consistent with the ones obtained by Wong {\it et al.} \cite{wong2012}. This verifies that the approximation we used in Eq. (\ref{tsallisus}) is good.
  
      \begin{figure*} 
        \centering
        \begin{tabular}{ccc}
        \includegraphics[width=0.35\textwidth]{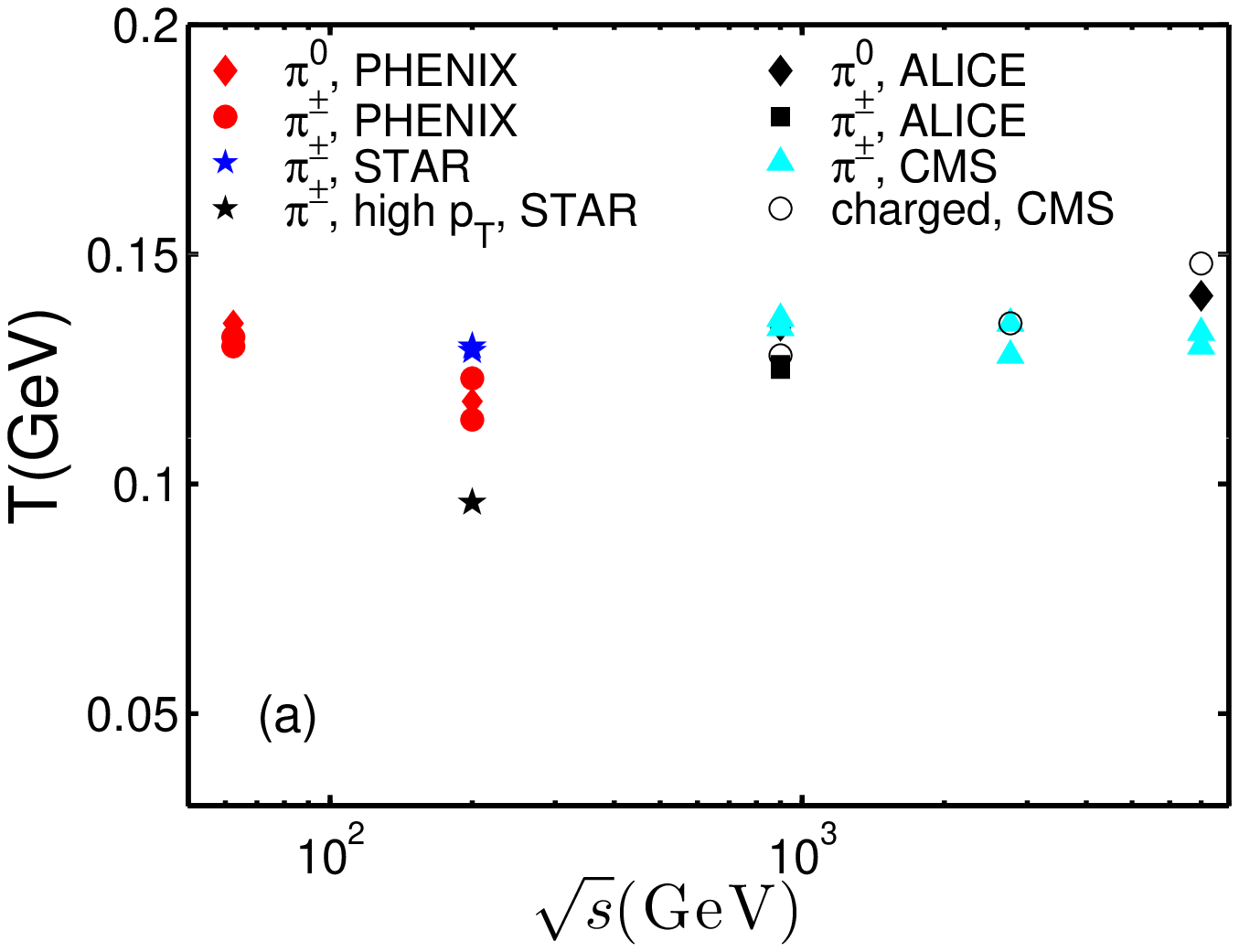}
        \includegraphics[width=0.35\textwidth]{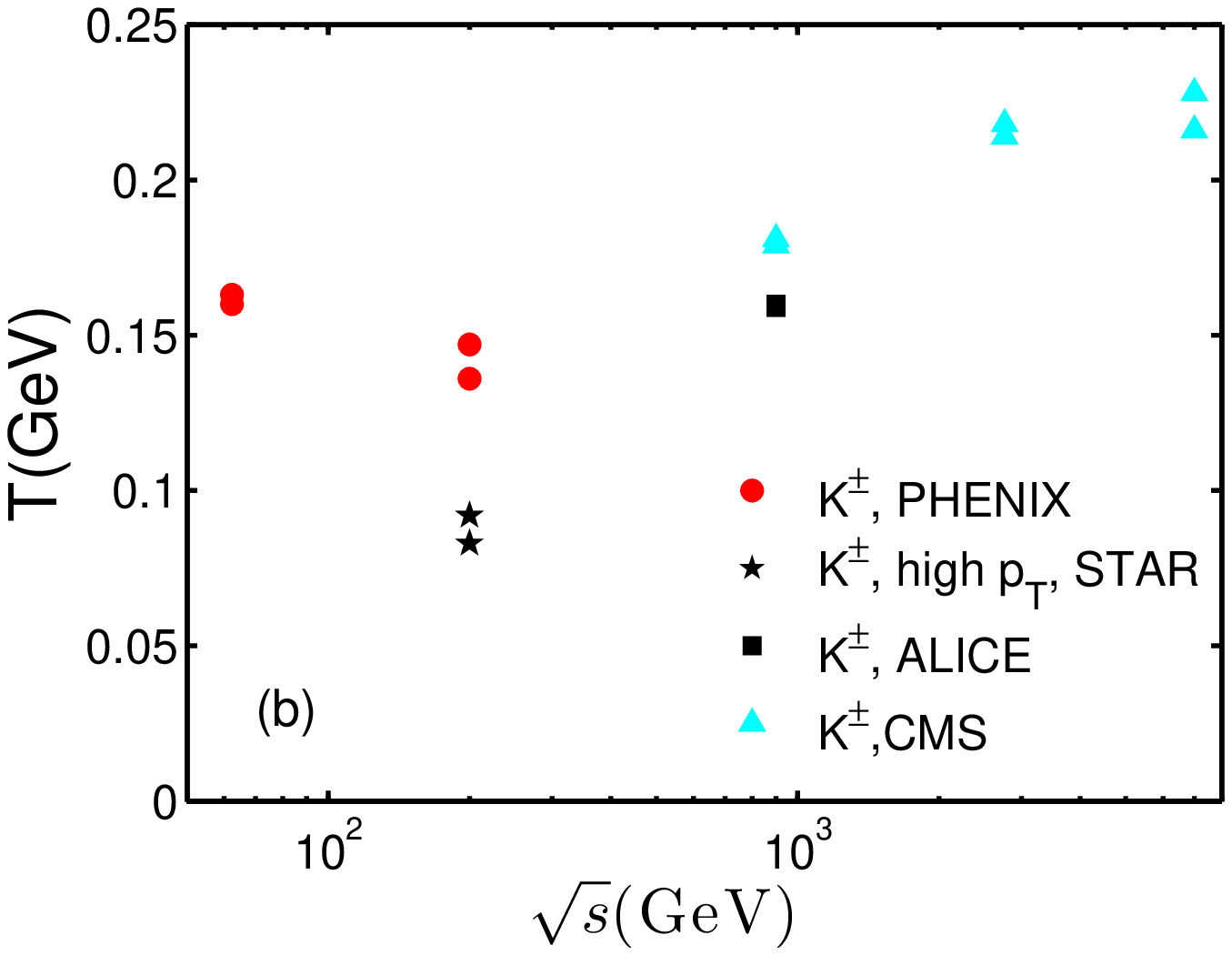}
          \includegraphics[width=0.35\textwidth]{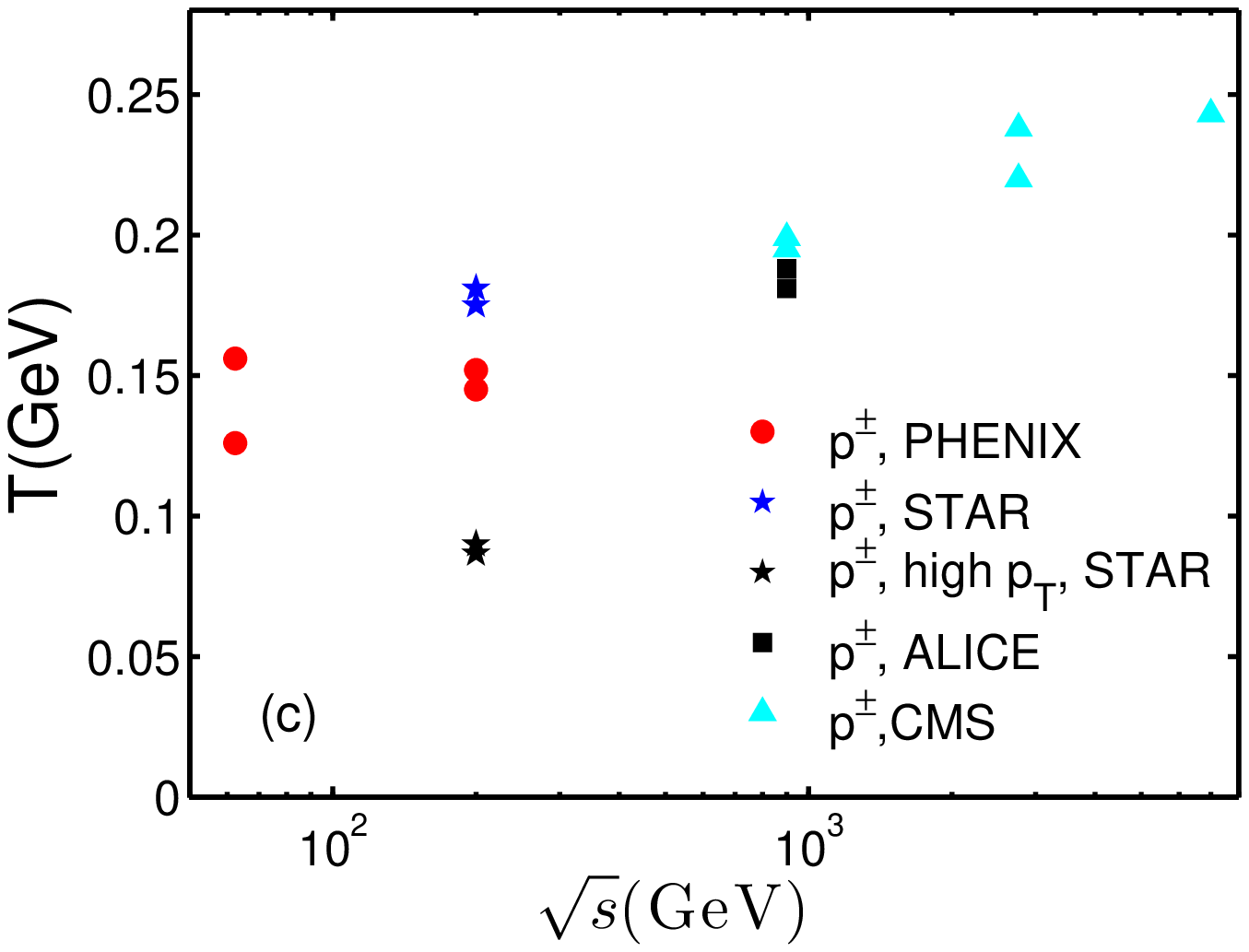}
               \end{tabular}
\caption{(Color online) Temperature $T$ in the Tsallis distribution versus $\sqrt{s}$ for (a) $\pi$, (b)kaon and (c) proton in p+p collisions. } \label{fitt}
    \end{figure*}
         \begin{figure*} 
        \centering
        \begin{tabular}{ccc}
        \includegraphics[width=0.35\textwidth]{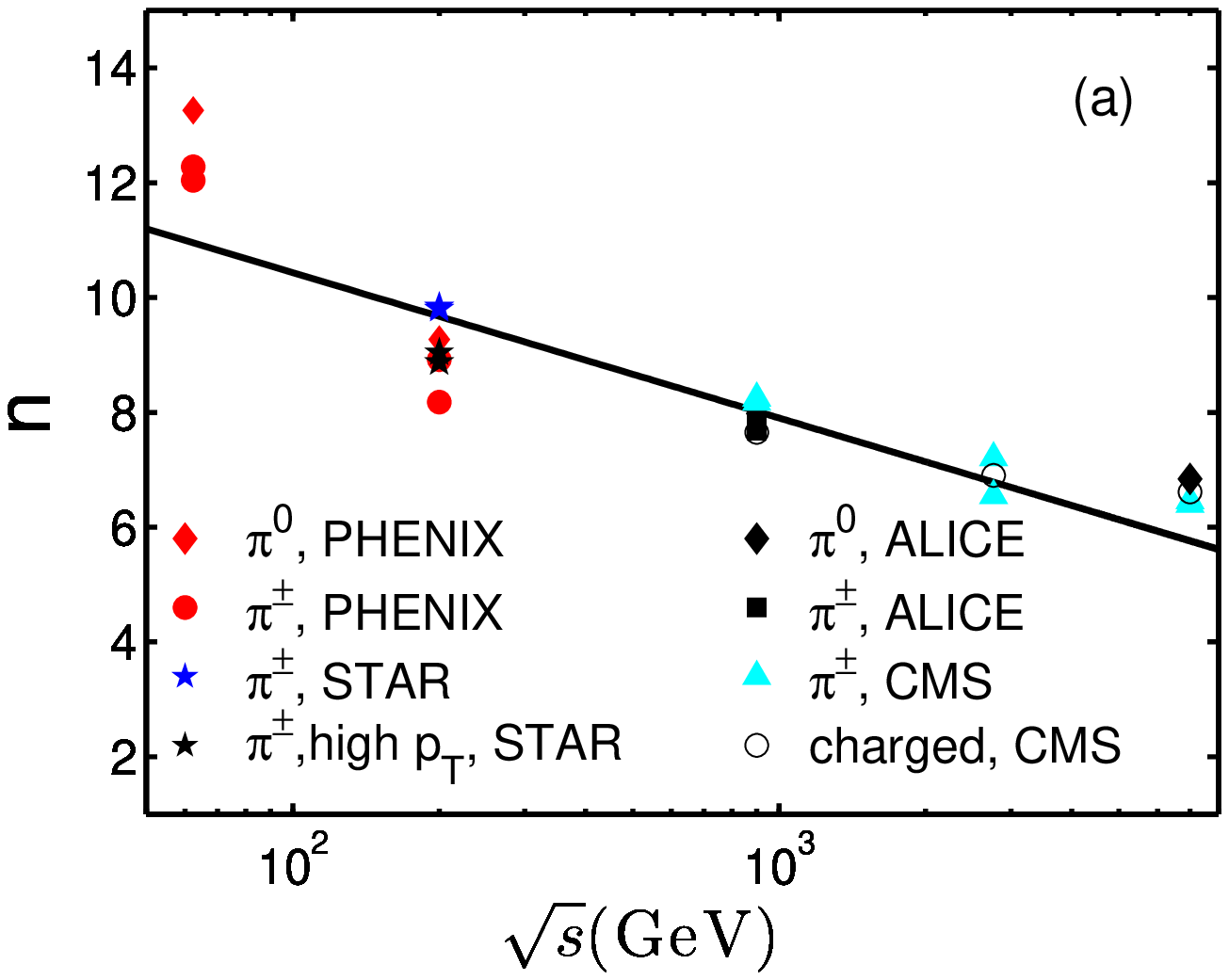}
        \includegraphics[width=0.35\textwidth]{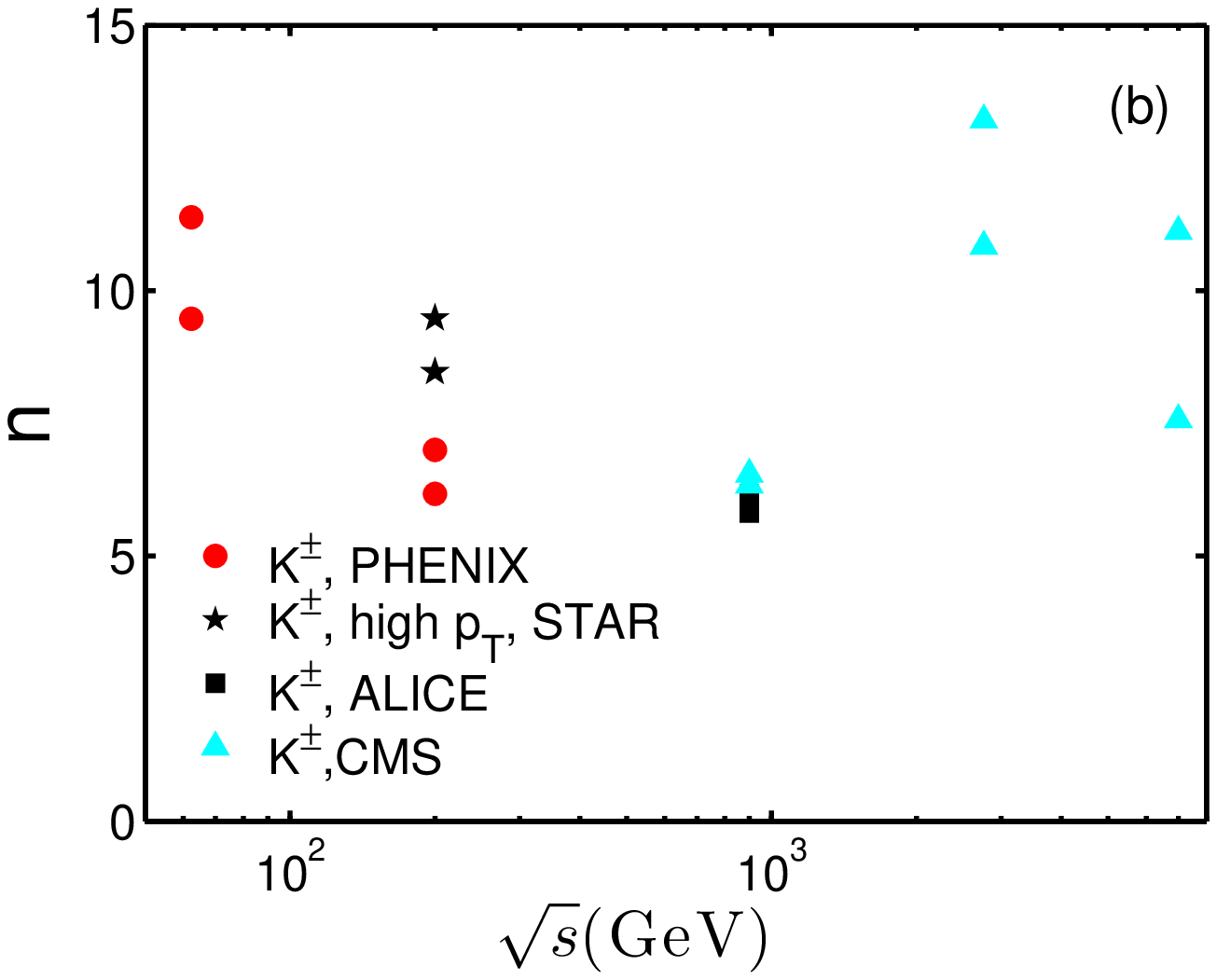}
          \includegraphics[width=0.35\textwidth]{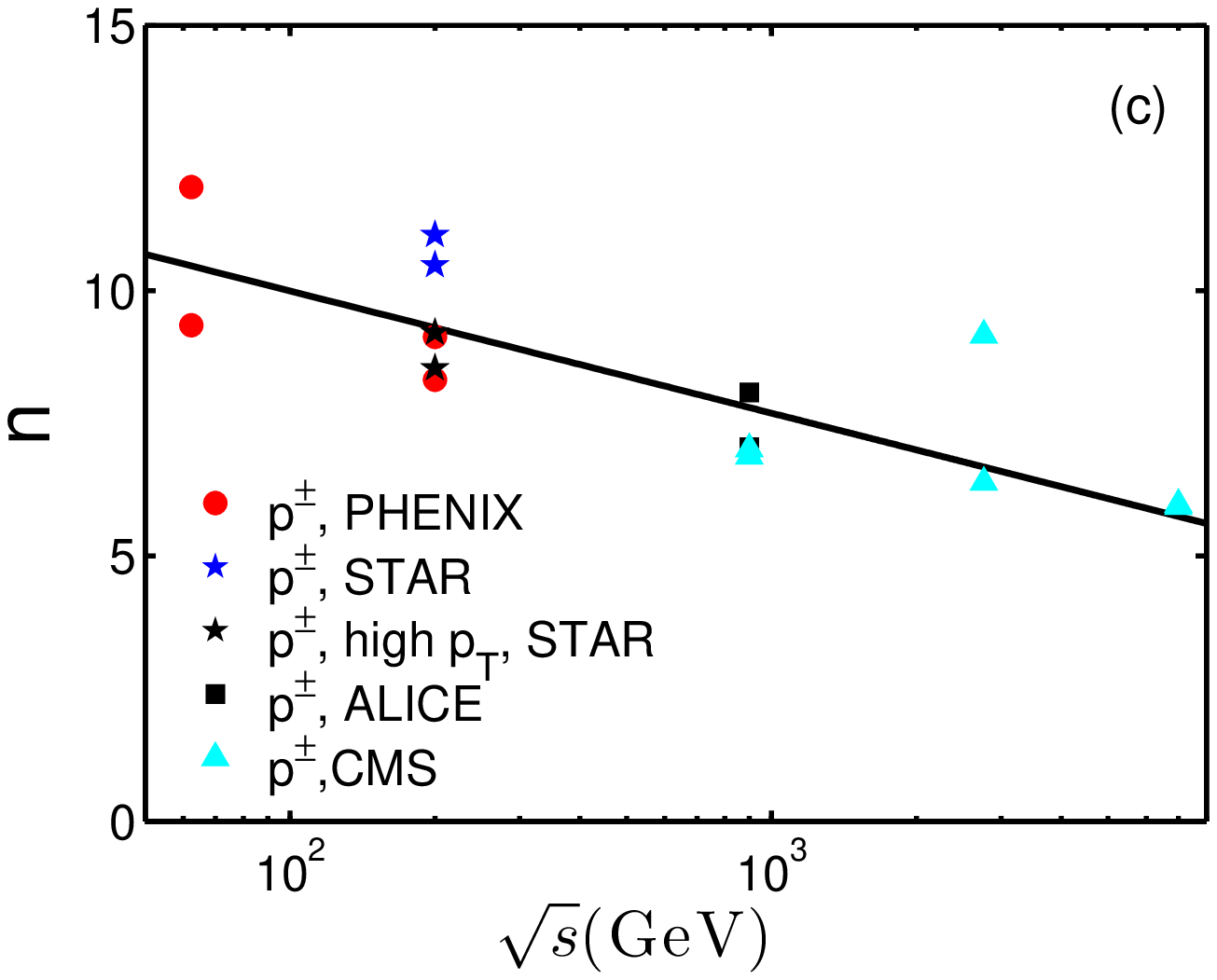}
               \end{tabular}
\caption{(Color online) The parameter $n$ of the Tsallis distribution versus $\sqrt{s}$ for (a) $\pi$, (b) kaon and (c) proton in p+p collisions. The lines are to guide the eyes.}\label{fitn}
    \end{figure*}
    
    In Fig. \ref{fitt}, we show the parameter $T$ of the Tsallis distribution obtained by fitting spectra of particles at different $\sqrt{s}$ in p+p collisions. Firstly, we can see that the parameter $T$ depends on the fitting $p_T$ range. Only the results for high $p_T$ particles (black stars) are very much different from the ones including all the $p_T$ particles at $\sqrt{s}=200$ GeV. Secondly, $T$ are almost the same for a particle and its anti-particle as it should be. Thirdly, $T$ for pions produced at different collision energy $\sqrt{s}$ is almost a constant around 0.13 GeV. This is consistent with the results in a similar study for charged particles in p+p collisions \cite{wongarxiv2014} and p+Pb collsions at $\sqrt{s_{NN}}=5.02$ TeV \cite{cmsppb1}. This may indicate that the pion production mechanism is similar in p+p and p+Pb collisions. $T$ for kaons is a little bit smaller than the one for protons at the same collision energy $\sqrt{s}$. But they have similar behaviors and increase with $\sqrt{s}$. We can interpret the results in the framework of a cascade particle production mechanism. Since the kaons and protons are heavier than pions, they are more likely to be produced at the beginning while pions can be produced at all times. Then their temperatures are higher than pions. When we include the results of the $\eta$ meson \cite{alice2} and the multi-strange baryons $\Xi^\pm$ and $\Omega^\pm$ \cite{aliceS2012} produced at $\sqrt{s}=7000$ GeV together, the trend is obvious. $T$ is higher when the mass of the particle is heavier which is consistent with the cascade particle production mechanism. Similar behaviors have been observed in p+Pb collisions as well \cite{cmsppb1}. Finally, $T$ for the identified hadrons depends on the Tsallis distribution form used in the fitting process. As we mentioned before, some authors argue that the identified hadrons should have the same $T$ \cite{khandai}; others suggest otherwise \cite{phenix2011, alice1, alice2, aliceS2012, cleymans, cmsppb1}. Therefore we only can compare the results of pions from different versions of the Tsallis distribution. In this case, we also can put the results of charged particles from different versions of the Tsallis distribution in since $T$ for charged particles is very similar to pions's as showed in the previous discussion. We categorize Eqs. (\ref{exptsallis}, \ref{tsalliswong}, \ref{tsallisus}) as type-A, Eqs. (\ref{tsallisB}, \ref{tsallisBR}) as type-B and Eqs. (\ref{senaeq}, \ref{tsallislong}) as type-C Tsallis distribution to clarify the discussion. Type-A Tsallis distribution gives $T\sim 0.13$ GeV for pions and charged particles. This has been shown in Fig. \ref{fitt} (left panel) and can be found in refs. \cite{phenix2011, alice1, alice2, wong2012, wongarxiv2014}. Type-B gives $T\sim 0.07$ GeV \cite{cleymans0, cleymans, azmiJPG2014, cleymans2}. The extra term $m_T$ in Eqs. (\ref{tsallisB}, \ref{tsallisBR}) is responsible for this lower $T$, which is the main difference between Type-A and Type-B. Without it, Type-B will give similar $T$ as Type-A. Type-C gives $T\sim 0.08$ GeV \cite{sena}. Another type of Tsallis distribution, see equation (7) in ref. \cite{khandai}, gives $T\sim 0.1$ GeV.

 As discussed in refs. \cite{khandai, wongprd}, we can connect the parameter $n$ in the power law of particle spectra at high $p_T$ with the particle production process. We refer to \cite{khandai, wongprd} for details. Here we just briefly state the results. The parameter $n=2(n_a-2)$ where $n_a$ is the number of active participants in the point like scattering. $n_a$ includes all the consituents of initial and final states in the scattering. If we assume that the basic scattering process at high $p_T$ is $qq\rightarrow qq$, we have 2 constituents for initial and final states respectively, then $n_a = 2+2 =4$ and $n=4$. We can obtain higher values of $n$ from the scattering at the hadron level, i.e. $p+meson\rightarrow p+meson$, we have $n_a = 5+5=10$ and $n=16$ which is the upper limit for $n$ for pions and kaons. For proton, $n$ can go higher up to 20 in the scattering process $pp\rightarrow pp$. This also sets the limits for the parameter $q$ in the Tsallis distribution. Using Eq. (\ref{nq}), we obtain $q\in [1.063, 1.25]$ for mesons and $q\in [1.05, 1.25]$ for protons. The fitting results of $q$ are between the limits in refs. \cite{sena, cleymans, azmiJPG2014}. In Fig. \ref{fitn}, we show the parameter $n$ obtained by fitting the spectra of particles with Tsallis distribution at different $\sqrt{s}$ in p+p collisions. We can see that the values of $n$ are well between those limits for all particles. This could give us some hints about the particle production mechanism in p+p collisions. We notice that the parameters $n$ for pions and protons are decreasing with the collision energy which indicates that interactions are from hadron level to quark level. This picture is quite clear we can break the hadrons into quarks at higher collision energy. But if we look at the results from kaons, the $n$ dereases first and then increases. Such a feature is not clear.


\begin{table}[H]
\begin{spacing}{0.7}
  \centering 
      \caption{The fitting parameters in Tsallis distribution Eq. (\ref{tsallisus}) for the particle spectra in p+p collisions.} 
 \resizebox{0.9\textwidth}{!}{%
            \begin{tabular}{c|c|c|c|c|c|c|c} 
\hline
data source                                 & $\sqrt{s}$(GeV) & particle & $p_T$(GeV)      & $|\eta|$     & $A$ & $n$ & $T$(GeV) \\
\hline
PHENIX\cite{phenix2009} &    62.4        & $\pi^0$   &  [0.614-6.734]   &  $<0.35$              & 0.128     & 13.26   & 0.135 \\
\hline
PHENIX\cite{phenix2011} &    62.4        & $\pi^+$   &  [0.35-2.85]   &  $<0.35$               & 0.104     & 12.04   & 0.132 \\
                 			    &                   & $\pi^- $   &  [0.35-2.85]   &                            & 0.111     & 12.28  & 0.130\\
                                              &                   & $K^+$    &  [0.45-1.95]   &                               & 0.0035  & 9.47    & 0.160\\
                                             &                    & $K^-$     &  [0.45-1.95]  &                                & 0.0031  & 11.38   & 0.163\\
                                             &                    & $p$        &  [0.65-3.5]    &                              & 0.0017  & 11.95   & 0.156 \\
                                             &                    & $\bar p$ & [0.65-3.5]     &                              & 0.0017  & 9.35     & 0.126 \\
\hline
PHENIX\cite{phenix2007} &    200     &   $\pi^0$   &  [0.616-18.93] & $<0.35$           & 0.218  &  9.27  & 0.118 \\

\hline
PHENIX\cite{phenix2011} &    200     &   $\pi^+$   &  [0.35-2.95] & $<0.35$             & 0.174  &  8.18  & 0.114 \\
                                             &                & $\pi^-$     &  [0.35-2.95]  &                           & 0.148  &  8.92  & 0.123\\
                                             &                & $K^+$     &  [0.45-1.95]  &                            & 0.0065 &6.17&0.136 \\
                                            &                 & $K^-$      &[0.45-1.95] &                         & 0.0058   &    7  &0.147 \\
                                            &                 & $p$         & [0.55-4.5]   &                         & 0.0029  & 8.32 & 0.145 \\
                                            &                 & $\bar p$  & [0.55-4.5] &                        & 0.0022 & 9.13 & 0.152 \\

\hline
STAR\cite{star2} &    200     &   $\pi^+$   &  [0.35-9.0] & $<0.5$ &            5.219 &  9.8  & 0.129 \\
                                             &                & $\pi^-$     &  [0.35-9.0]  &               &  5.077  &  9.84  & 0.130\\
                                            &                 & $p$         & [0.468-6.5]   &                          & 0.073  & 11.05 & 0.181 \\
                                            &                 & $\bar p$  & [0.468-6.5] &                         & 0.0629 & 10.48 & 0.175 \\
                                            
\hline
STAR\cite{star1} &    200     &   $\pi^+$   &  [3.113-13.06] & $<0.5$ &             20.809  &  8.87  & 0.096 \\
                                             &                & $\pi^-$     &  [3.113-13.06]  &                         & 21.557  &  9.04  & 0.096\\
                                             &                & $K^+$     &  [3.113-13.06]  &                           & 7.666 &8.47&0.083 \\
                                            &                 & $K^-$      &[3.113-13.06] &                       & 6.123   &    9.48  &0.092 \\
                                            &                 & $p$         & [3.113-13.06]   &                           & 2.2782  & 8.54 & 0.0867 \\
                                            &                 & $\bar p$  & [3.113-13.06] &                         & 2.2923 & 9.21 & 0.09 \\
     
\hline
ALICE)\cite{alice1}    &   900  & $\pi^+$  & [0.11-2.5] &     $<0.9$ &                 5.333      &   7.68    &  0.125       \\  
                                     &          & $\pi^-$   &[0.11-2.5]  &                 &                    5.279      &  7.85    & 0.126   \\
                                    &          & $K^+$    &[0.225-2.3] &               &                0.206   &  5.81    & 0.159 \\
                                   &           & $K^-$    &[0.225-2.3] &               &                 0.208   & 6.17     & 0.160 \\                             
                                    &          & $p$       & [0.375-2.3] &               &               0.0524   & 7.05    & 0.181 \\  
                                    &          & $\bar p$& [0.375-2.3] &               &                0.0498   & 8.08   & 0.188 \\     
\hline
ALICE\cite{alice2}   &  900   & $\pi^0$  & [0.495-5.818]&  0        &                  0.26    &  7.98   &  0.135  \\                               
\hline
CMS\cite{cms3}    &   900  & $\pi^+$  & [0.125-1.175] &     $<1.0$ &                    6.139      &   8.25    &  0.134       \\  
                                     &          & $\pi^-$   & [0.125-1.175]  &                 &                     5.827      &  8.18    & 0.136   \\
                                    &          & $K^+$    &[0.225-1.025] &               &                 0.237   &  6.54    & 0.181 \\
                                   &           & $K^-$    &[0.225-1.025] &               &                  0.235   & 6.34     & 0.179 \\                             
                                    &          & $p$       & [0.375-1.675] &               &                0.0618   & 7.02    & 0.199 \\  
                                    &          & $\bar p$& [0.375-1.675] &               &                0.0598   & 6.88   & 0.195 \\                                               
\hline
CMS\cite{cms3}    &   2760  & $\pi^+$  & [0.125-1.175] &     $<1.0$ &                   7.702      &   6.55    &  0.128       \\  
                                     &          & $\pi^-$   & [0.125-1.175]  &                 &                    7.11      &  7.21    & 0.135   \\
                                    &          & $K^+$    &[0.225-1.025] &               &                 0.264   &  10.84    & 0.214 \\
                                   &           & $K^-$    &[0.225-1.025] &               &                  0.258   & 13.22    & 0.218 \\                             
                                    &          & $p$       & [0.375-1.675] &               &                0.0678   & 6.39    & 0.22 \\  
                                    &          & $\bar p$& [0.375-1.675] &               &                0.0632   & 9.16   & 0.238 \\                                                  
\hline
 ALICE\cite{alice2}   &   7000  & $\pi^0$  & [0.35-22.197] &     0 &                  0.467    &   6.84    &  0.141       \\   
\hline
CMS\cite{cms3}    &   7000  & $\pi^+$  & [0.125-1.175] &     $<1.0$ &                    9.298     &   6.4    &  0.130       \\  
                                     &          & $\pi^-$   & [0.125-1.175]  &                 &                    8.815      &  6.46    & 0.133   \\
                                    &          & $K^+$    &[0.225-1.025] &               &                0.302   &  11.12   & 0.228 \\
                                   &           & $K^-$    &[0.225-1.025] &               &                 0.307   & 7.57    & 0.216 \\                             
                                    &          & $p$       & [0.375-1.675] &               &               0.0742   & 5.93   & 0.243 \\  
                                    &          & $\bar p$& [0.375-1.675] &               &               0.074   & 5.97 & 0.243 \\  
\hline             
CMS\cite{cms1}   &  900           & charged   &   [0.5-31.2]      &     $<2.4$                 &  15.938     &  7.65    &  0.128  \\  
CMS\cite{cms2}   &  2760         & charged   &   [0.525-99.3]  &      $<1$                   &  17.661     &  6.9      &  0.135  \\   
CMS\cite{cms1}   &  7000         & charged   &   [0.5-181.2]    &     $<2.4$                  &  15.92       &  6.61    &  0.148  \\                                      \hline

  \end{tabular}
} 
         \label{fitpara}
\end{spacing}
\end{table}

\section{Conclusions}
We have made a thorough study of traverse momentum spectra of identified particles produced in p+p collisions at RHIC and LHC energies with Tsallis distribution. A detailed analysis of the parameters $T$ and $n$ is also shown.  $T$ is not dependent on the beam energy for pions, while for kaons and protons it increases with increasing energy. Furthermore, we notice that $T$ is higher for the particle whose mass is larger. This is probably due to the particle produced time. In the cascade particle production mechanism, this result is perfectly understandable. The behavior of $n$ for kaons is not the same as pions and protons, which is related to the particle production process. From the properties of $T$ and $n$, we get more information about particle production mechanism in p+p collision. We also wish more exciting results can be found in p+p collisions at 8 TeV.  

\section*{Acknowledgments} 
We thank Prof. J. Natowitz for discussions. This work is supported by the NSFC of China under Grant No.\ 11205106.


\begin{thebibliography}{99}

\bibitem{star2007} 
B.I. Abelev {\it et. al.} (STAR Collaborations), \Journal{\PRC}{75}{064901}{2007} 

\bibitem{phenix2011}
A. Adare {\it et al.} (PHENIX Collaboration), \Journal{\PRC}{83}{064903}{2011} 

\bibitem{alice1} 
K. Aamodt {\it et al.} (ALICE Collaboration), \Journal{\EPJC}{71}{1655}{2011}

\bibitem{alice2} 
B. Abelev {\it et al.} (ALICE Collaboration), \Journal{\PLB}{717}{162}{2012} 

\bibitem{aliceS2012} B. Abelev {\it et al.} (ALICE Collaboration), \Journal{\PLB}{712}{309}{2012}

\bibitem{cms3} S. Chatrchyan {\it et al.} (CMS Collaboration), \Journal{\EPJC}{72}{2164}{2012}

\bibitem{cmsdata900} V. Khachatryan {\it et al.} (CMS Collaboraction), \Journal{\JHEP}{02}{041}{2010} 

\bibitem{cmsdata7000} V. Khachatryan {\it et al.} (CMS Collaboraction), \Journal{\PRL}{105}{022002}{2010} 

\bibitem{cms1} S. Chatrchyan {\it et al.} (CMS Collaboration),  \Journal{\JHEP}{08}{086}{2011} 

\bibitem{sena} I. Sena and A. Deppman, \Journal{\EPJA}{49}{17}{2013} 

\bibitem{liuAuAu2014}F.H. Liu, Y.Q. Gao and B.C. Li, \Journal{\EPJA}{50}{123}{2014}

\bibitem{khandaiflow}P.K. Khandai, P. Sett, P. Shukla and V. Singh, \Journal{\JPG}{41}{025105}{2014}

\bibitem{cms2014}  S. Chatrchyan {\it et al.} (CMS Collaboration), \Journal{\EPJC}{74}{2847}{2014}

\bibitem{cleymans0}L. Marques, J. Cleymans and A. Deppman, \Journal{\PRD}{91}{054025}{2015} 

\bibitem{cleymans}J. Cleymans and D. Worku, \Journal{\JPG}{39}{025006}{2012} 

\bibitem{azmiJPG2014} M.D. Azmi and J. Cleymans, \Journal{\JPG}{41}{065001}{2014} 

\bibitem{liAuAu2013}B.C. Li, Y.Z. Wang and F.H. Liu, \Journal{\PLB}{725}{352}{2013}

\bibitem{maciej} M. Rybczy\'nski and Z. W\l odarczyk, \Journal{\EPJC}{74}{2785}{2014} 

\bibitem{khandai} P.K. Khandai, P. Sett, P. Shukla and V. Singh, \Journal{\IJMPA}{Vol. 28, No. 16}{1350066}{2013}

\bibitem{wongprd}Cheuk-Yin Wong, and G. Wilk, \Journal{\PRD}{87}{114007}{2013}

\bibitem{wong2012}Cheuk-Yin Wong, and G. Wilk, Acta Physica Polonica B, {\bf Vol.43}, No. 11, (2012).


\bibitem{wongarxiv2014}Cheuk-Yin Wong, G. Wilk, L. J. L. Cirto and C. Tsallis, \Journal{\PRD}{91}{114027}{2015}

\bibitem{maciej1} G. Wilk and Z. W\l odarczyk, arXiv: 1503.08704v1.

\bibitem{cleymans2} J. Cleymans, G.I. Lykasov, A.S. Parvan, A.S. Sorin, O.V. Teryaev and D. Worku, Phys. Lett. B {\bf723}, 351(2013)

\bibitem{wongbook} Cheuk-Yin Wong, {\em Introduction to High-Energy Heavy-Ion Collisions,} (World Scientific, Singapore, 1994)

\bibitem{beck} C. Beck, \Journal{\PA}{286}{164}{2000}

\bibitem{dndeta} I. Bautista, C. Pajares and J. Dias de Deus, \Journal{\NPA}{882}{44}{2012}

\bibitem{cmsB2013} S. Chatrchyan {\it et al.} (CMS Collaboration), \Journal{\PLB}{714}{136}{2012}

\bibitem{star2010} B.I. Abelev {\it et al.} (STAR Collaboration), \Journal{\PRC}{81}{064904}{2010}

\bibitem{phenix2009} A. Adare {\it et al.} (PHENIX Collaboration), \Journal{\PRD}{79}{012003}{2009}

\bibitem{phenix2007}  A. Adare {\it et al.} (PHENIX Collaboration), \Journal{\PRD}{76}{051106(R)}{2007}

\bibitem{star2} J. Adams {\it et al.} (STAR Collaboration), \Journal{\PLB}{637}{161}{2006}

\bibitem{star1} G. Agakishiev {\it et al.} (STAR Collaboration), \Journal{\PRL}{108}{072302}{2012}

\bibitem{cms2} S. Chatrchyan {\it et al.} (CMS Collaboration), \Journal{\EPJC}{72}{1945}{2012}

\bibitem{cmsppb1} CMS Collaboration, \Journal{\EPJC}{74}{2847}{2014}

\end{thebibliography}
\end{document}